\documentclass[]{piparticle-final}
\usepackage{graphicx}
\usepackage{amsmath}

\usepackage{cite} 
\usepackage{epstopdf} 

\begin{document}
\volume{8}               
\articlenumber{080006}   
\journalyear{2016}       
\editor{K. Hallberg}   
\received{{2} August 2016}     
\accepted{12 October 2016}   
\runningauthor{D. Dey\itshape{et al.}}  
\doi{080006}         

\title{An efficient density matrix renormalization group  algorithm for chains with periodic boundary condition}

\author{Dayasindhu Dey,\cite{inst1}
        Debasmita Maiti,\cite{inst1}
        Manoranjan Kumar\cite{inst1}\thanks{E-mail: manoranjan.kumar@bose.res.in}}

\pipabstract{
The Density Matrix Renormalization Group (DMRG) is a state-of-the-art numerical technique for a one dimensional quantum 
many-body system; but calculating accurate results for a system 
with Periodic Boundary Condition (PBC) from the conventional DMRG has 
been a challenging job from the inception of DMRG. The recent 
development of the Matrix Product State (MPS) algorithm gives 
a new approach to find accurate results for the one dimensional PBC system. 
The most efficient implementation of the MPS algorithm can scale as 
O($p \times m^3$), where $p$ can vary from 4 to $m^2$. 
In this paper, we propose a new DMRG algorithm, 
which is very similar to the conventional DMRG and gives comparable 
accuracy to that of MPS. The computation effort of the new algorithm 
goes as O($m^3$) and the conventional DMRG code can be easily modified for 
the new algorithm.
}

\maketitle

\blfootnote{
\begin{theaffiliation}{99}
   \institution{inst1} S. N. Bose National Centre for Basic Sciences, Block JD, Sector III, Salt Lake, Kolkata 700106, India.
\end{theaffiliation}
}

\section{Introduction}
The quantum many-body effect in the condensed matter gives 
rise to many exotic states such as superconductivity~\cite{tinkham2004}, 
multipolar phases~\cite{chubkov91,multimagnon-prb2007}, 
valence bond state~\cite{anderson-vb87}, vector 
chiral phase~\cite{chubkov91,aslam-jmmm} and topological superconductivity~\cite{tis-rev-rmp2011}. 
These effects are prominent in the one dimensional (1D) electronic systems
due to the confinement of electrons.  The confinement of electrons 
and the competition between the electron-electron repulsion and the kinetic energies of 
electrons produce many interesting phases like Spin Density Wave (SDW), 
dimer  or the bond order wave phase and Charge Density Wave (CDW) phase 
in one dimensional systems~\cite{nakamura2000,pinanki2002,mk2009}. Although these 
quantum many-body effects in the system are crucial for exotic phases, dealing 
with these systems is a challenging job because of the large degrees of 
freedom. The degrees of freedom increase as $2^N$ or $4^N$ for a 
spin-1/2 system or a fermionic system, respectively.

In most cases, the exact solutions for these systems with large degrees of freedom are almost impossible to reach. Therefore, during the last 
three decades many  numerical techniques have been developed, e.g., Quantum 
Monte Carlo (QMC)~\cite{qmc-suzuki}, Density Functional Theory (DFT)~\cite{kohn65}, 
Renormalization Group (RG)~\cite{wilson83} and Density Matrix Renormalization Group (DMRG) method~\cite{white-prl92,white-prb93}. The DMRG is a  
state-of-the-art numerical technique for 1D systems with Open Boundary Condition (OBC). 
However, the numerical effort to maintain the accuracy for PBC systems 
becomes exponential~\cite{schollwock2005,hallberg2006}. It is 
well known that the Periodic Boundary Condition (PBC) is essential to get rid 
of the boundary effect of a finite open chain and also to preserve the inversion 
symmetry in the systems~\cite{soos-jpcm-2016}.

The DMRG technique is based on the systematic truncation of irrelevant degrees 
of freedom and has been reviewed extensively in  Ref.~\cite{schollwock2005,hallberg2006}. 
In a 1D system with OBC, the number of relevant degrees of freedom is 
small~\cite{schollwock2005,hallberg2006}. Let us consider that for a given 
accuracy of the OBC system, $m_{\rm_{obc}}$ number of eigenvectors 
of the density matrix is required, then the conventional DMRG for the PBC 
system  requires O($m^2_{\rm{obc}}$)~\cite{Schollwock2011}. In the 
conventional DMRG, computational effort for the OBC systems with sparse 
matrices goes as O($m^3_{\rm{obc}}$), whereas, it goes as O($m^6_{\rm{obc}}$) 
for the PBC system~\cite{pippan2010}. The accuracy of energies for 
the PBC systems calculated from the conventional DMRG decreases 
significantly, and there is a long bond in the system which connects
both ends. 

The accuracy of operators decreases with the number of
renormalization, especially the raising/creation $S^+/a+$ and lowering/annihilation $S^-/a^-$ 
operator of spin/fermionic systems. The conventional DMRG is solved in a 
$S^z$ basis, therefore the exact $S^z$ operator remains diagonal and, 
multiple times, renormalization deteriorates the accuracy slowly; 
but   $S^+/a+$ and $S^-/a^-$ are off diagonal in this exact basis and 
therefore, the multiple time renormalization of these operators decreases 
the accuracy of the operators.  A similar type of accuracy problems occurs 
for multiple time renormalized $ a^+$ and $a^-$ in the fermionic systems. 
In fact, it has been noted that accuracy of energy of a system with PBC 
significantly increases if the superblock is constructed with very few 
times renormalized operators~\cite{mk2009}. To avoid multiple renormalization,
new sites are added at both ends of the chain in such a way that 
only second time renormalized operators are used to construct the
superblock. In this algorithm, there is a connectivity between the old-old
sites and their operators are renormalized; and this connectivity spoils 
the sparsity of the superblock Hamiltonian~\cite{mk2009}.

In this paper, a new DMRG algorithm  is proposed, which can be implemented upon the  existing 
conventional DMRG code in a few hours and gives  accurate results which are comparable 
to those of MPS algorithm.  In fact, this algorithm can be implemented for two-legged
ladders without much effort~\cite{dagotto-science96}. We have studied the spectrum of 
density matrix of the system block, ground state energy and correlation 
functions of a Heisenberg anti-ferromagnetic Hamiltonian for spin-1/2 and spin-1 on a 1D chain 
with PBC.  

This paper is divided into five sections. The model Hamiltonian is 
discussed in section~II. The new algorithm and the comparative studies 
of algorithms are done in section~III. The accuracy of various quantities 
is studied in the section~IV. In section~V, results and algorithm are discussed.

\section{\label{sec:model}Model Hamiltonian}

Let us consider a strongly correlated electronic system where Coulomb 
repulsion is dominant, therefore the charge degree of freedom gets 
localized, for example, Hubbard model in large $U$ limit in a half 
filled band. In this limit, the system becomes insulating, but the  electrons 
can still exchange their spin. The Heisenberg model is one of the most 
celebrated models in this limit, and only the spin degrees of freedom are active
in the model. The Heisenberg model Hamiltonian can be written as 

\begin{align}
H = \sum_{\langle i j \rangle} J_{ij} \vec{S}_i \cdot \vec{S}_j
\end{align}

where, $J_{ij}$ is the anti-ferromagnetic exchange interaction 
between the nearest neighbor spin. In the rest of the paper, $J_{ij}$ 
is set to be 1.

\section{\label{sec:alg}Comparison of algorithms}

A ground state wavefunction calculated from the conventional DMRG 
can be represented in terms of the Matrix Product State (MPS), 
as shown by Ostlund and Rommer~\cite{ostlund-rommer95}. The 
wavefunction can be written as

\begin{align}
|\psi \rangle = \sum_{n_1,n_2,\ldots,n_L} {\rm{tr}} (A_{n_1}^1 A_{n_2}^2 \ldots
A_{n_L}^L) | n_1,n_2,\ldots,n_L \rangle,
\end{align}
where $A_{n_k}^k$ are a set of matrices of dimension $m \times m$ for site $k$
and with $n_k$ degrees of freedom. The wave function $|\psi \rangle$ 
can be accurately found if $m$ is sufficiently large. The expectation 
value of an operator $O_k$ in the gs~\cite{pippan2010,verstraete-prl04} can 
be written as

\begin{align}
\langle O_k \rangle = {\rm{tr}} \left( \sum_{n_k, n^{'}_k} \langle n_k|O_k|n^{'}_k
\rangle A_{n_k}^k \otimes A_{n^{'}_k}^k \right),
\end{align}
where $n_k$ is the local degrees of freedom of site $k$.
The matrix $A$ can be evaluated by using the following equations

\begin{align}
H_k |\psi_k \rangle &= \lambda N_k |\psi_k \rangle, \\
N^k &= E_1^{k+1} E_1^k \ldots E_1^L E_1^{1} \ldots E_1^{k-1},
\end{align}
where

\begin{align}
E_1^k = \sum_{n_k, n^{'}_k} \langle n_k|1|n^{'}_k \rangle 
A_{n_k}^k \otimes A_{n^{'}_k}^k.
\label{eq:e-def}
\end{align}
Here, $H_k$ is the effective Hamiltonian of $k^{\rm{th}}$ site 
and $\lambda$ is the expectation value of energy.  The $A$ 
matrices are evaluated at this point and the matrices are rearranged
to keep the algorithm stable.  The $H_k$ and $N$ can be calculated 
recursively while evaluating $A$, one site at a time~\cite{pippan2010}.  
Here, $N$ matrices are ill-conditioned and require storing, 
approximately $L^2$ matrices as well as  multiplication of $L^2$ 
matrices of $m \times m$ size~\cite{pippan2010} at each step. 
The evaluation of $A$ and $N$ are done for all the sites and 
backward and forward sweeps for all the sites are executed similarly to the finite system DMRG. 
The mathematical operations of matrices of dimension $m^2 \times m^2$ 
 represent the Hamiltonian cost $\sim (o)m^6$, but the special 
form of these matrices reduces the cost by a factor of $m$. Therefore, this 
algorithm scales as $\sim (o) m^5$~\cite{pippan2010}.

The above algorithm is extended by Verstraete et. al., for translational 
invariant systems~\cite{verstraete2005}. Only two types of matrices, $A^1$ and $A^2$, 
are considered~\cite{verstraete2005}. Product of the two matrices can be 
repeated to compute $N$. In this algorithm, only two matrices, 
$A^1$ and $A^2$, are updated and optimized to get the gs properties. 
This algorithm scales as (o)$m^3$, although it does not work for a 
finite system, or systems with impurity, etc. Pippan et. al. 
introduced another MPS based efficient algorithm for translational
invariant PBC systems~\cite{pippan2010}.  In the old version of MPS, most of
the computation cost goes to constructing the product of $m^2 \times m^2$
matrices $E$ defined in  Eq. (\ref{eq:e-def}).  
The new MPS algorithm overcomes this problem by performing a SVD of the
product of sufficiently large $(l \gg 1)$ number of $m^2 \times m^2$ 
transfer matrices\cite{pippan2010,rossini2011a}. The singular values,
in general, decay very fast; therefore, only $p \, (\ll m^2)$ among 
$m^2$ singular values are significant~\cite{rossini2011a}. Thus, the 
computational cost now is reduced to (o) $p \times m^3$~\cite{pippan2010}.
However, one requires $p \sim m$ to reach adequate numerical accuracy 
of physical measures, as pointed out in Ref.~\cite{rossini2011a}.

Although the above technique is efficient and accurate, there 
are various reasons for developing the new algorithm. First, the
modified MPS works efficiently for a system where the singular values of 
products of matrices decay exponentially and this algorithm scales 
as (o) $p m^3$, where $p$ can vary linearly with $m$. Second, the 
implementation of the MPS based numerical technique is quite 
different from the conventional DMRG, and the MPS algorithm should be 
written from scratch. Third, many conventional numerical techniques 
like the dynamical correction vector~\cite{ramasesha-prb99} or continued 
fraction~\cite{cont-frac-hallberg}, and implementation of symmetries like parity or inversion 
symmetries are difficult. In this paper, we will explain a new 
algorithm which is very similar to the conventional DMRG technique, 
and also show that the new algorithm can give accuracy comparable to
that of MPS based techniques. This algorithm is applied for 
$S=1/2$ and $S=1$ chains with PBC. But first, let us try 
to understand the algorithm before discussing the results.

In this algorithm, we will try to avoid the multiple 
renormalization of operators, whereas the other parts of the 
algorithm remain the same as the conventional DMRG.
Before going to the new algorithm, let us recap the conventional DMRG.
\begin{enumerate}
\item Start with a superblock of four sites consisting of one 
site for both the left and the right block and two new sites.

\item Get desired eigenvalues and eigenvectors of the superblock and 
construct the density matrix $\rho$ of the system which consists of the 
left or the right block and a new site.

\item Now, construct an effective $\tilde{\rho}$  with $m$ number of 
eigenvectors of $\rho$, corresponding to the $m$ largest eigenvalues. 
The effective system Hamiltonian and  all operators in the 
truncated basis are constructed using the following equations:

\begin{align}
H = \tilde{\rho}^{\dagger} H \tilde{\rho}; \qquad 
O = \tilde{\rho}^{\dagger} H \tilde{\rho}
\label{eq:renorm}
\end{align}

\item Superblock is constructed using the effective Hamiltonian 
and operators of the system block and two new sites.

\item Repeat all the steps from 2 until the desired system size is reached. 
The full process is called infinite DMRG.
\end{enumerate}

As mentioned earlier, the conventional algorithm is excellent 
for a 1D open chain as superblock is constructed with only one time 
renormalized operators. However, for a PBC system, one needs a 
long bond; therefore, at least two operators of superblocks 
are renormalized multiple times. In the new algorithm, the multiple 
time renormalization of operator is avoided and the algorithm 
goes as: 

\begin{enumerate}
\item Start with a superblock with four blocks consisting of a 
left and a right block and two new site blocks. The blocks are 
shown in Fig.~\ref{fig1} as filled circles and may have more than one 
site. Here, we have considered only one site in each block. 
New blocks may also have more than one site and are shown as open 
circles. In this paper, new blocks have one site in a chain or  
two for a ladder, like a structure with PBC~\cite{soos-jpcm-2016}.   

\begin{figure}
\centering
\includegraphics[width=\columnwidth]{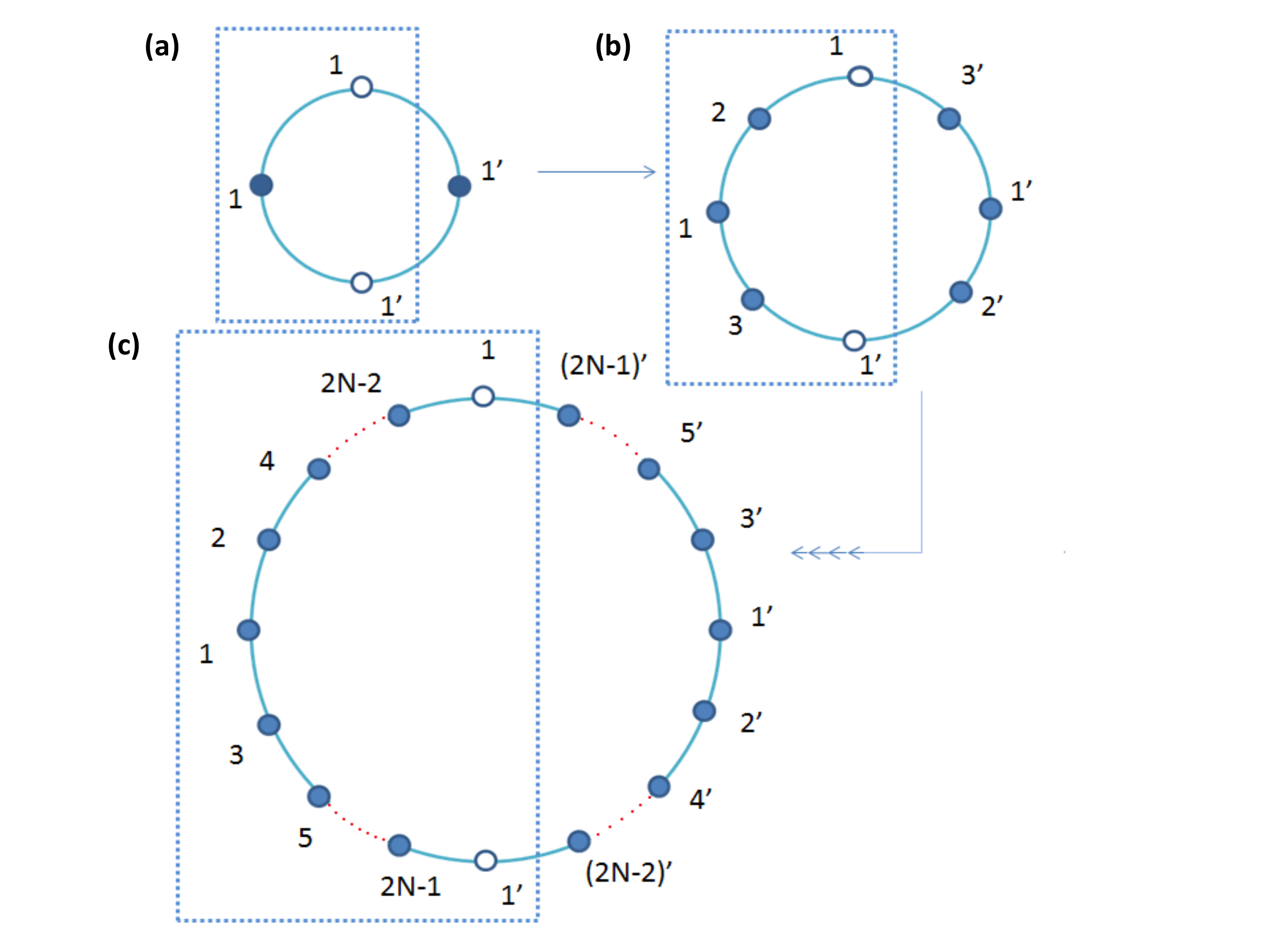}
\caption{\label{fig1}Pictorial representation of  the new 
DMRG algorithm with only one site in the new block. (a) One starts 
with two blocks left and right represented by filled circles and 
two new sites blocks represented as open circles.  The dotted 
box represents the system block for the next step. (b) Superblock of 
the next DMRG step is shown. (c)  The final step of infinite DMRG  
of $N=4N$ system size is shown with $2N-1$ number of sites in each of
the left and the right blocks and two new sites.}
\end{figure}

\item Get the eigenvalues and eigenvectors of the superblock and 
construct the density matrix $\rho$ of the system which 
consists of the left or the right block and two new blocks. 
The left system block is shown inside the box in Fig.~\ref{fig1}a.
   
\item Now, construct an effective $\tilde{\rho}$  with $m$ 
number of eigenvectors of $\rho$ corresponding to the $m$ 
largest eigenvalues of the density matrix.  The effective 
system Hamiltonian and operators in the truncated basis 
are calculated using Eq.~(\ref{eq:renorm}).
 
\item The superblock is constructed using the effective 
Hamiltonian and operators of system blocks and two new sites.
  
\item Now, go to step 2 and the processes 2 -- 5 are repeated 
till the desired system size is reached. 
\end{enumerate}

We notice that the superblock Hamiltonian is constructed using 
the effective Hamiltonian of blocks and operator which are 
renormalized once. Therefore, the massive truncation because of 
long bond is avoided in this algorithm.  Bonds in the superblock 
are only between new-new sites or new-old sites. For the construction 
of a Hamiltonian matrix of old-new site bond, a new site operator is 
highly sparse. However, old sites renormalized operators are highly 
dense. The old-old sites interaction in the conventional algorithm 
generates a large number of non-zero matrix elements in the superblock 
Hamiltonian matrix and the diagonalization of dense matrix goes as $m^4$.
But, in the new algorithm, old-new sites interaction in the superblock 
generates only a sparse Hamiltonian matrix, and its digonalization 
scales as $m^3$.

\section{\label{sec:res}Results}

We consider spin-1/2 and spin-1 chains with PBC of length up to $N=500$
to check the accuracy of results for the Heisenberg 
Hamiltonian.  In this part, we study the truncation error of 
density matrix $T$,  error in relative ground state energy  
$\frac{\Delta E}{|E_0|}$ and dependence of correlation function 
$C(r)$ on $m$. The correlation function $C(r)$ is defined as

\begin{align}
C(r) = \vec{S}_0 \cdot \vec{S}_r,
\label{eq:crdef}
\end{align}
where $\vec{S}_0$ corresponds to the reference spin and $\vec{S}_r$ is
the spin at a distance $r$ from the reference spin.
The relative ground state energy  can be defined as 
$\Delta E /|E_0|$, where  $\Delta E = E(m) - E_0$ with 
$E_0$ is the  most accurate value for $S=1$ chain~\cite{pippan2010} and 
$E_0=E(m=1200)$ for $S=1/2$ chain.  
\begin{figure}
\centering
\includegraphics[width=\columnwidth]{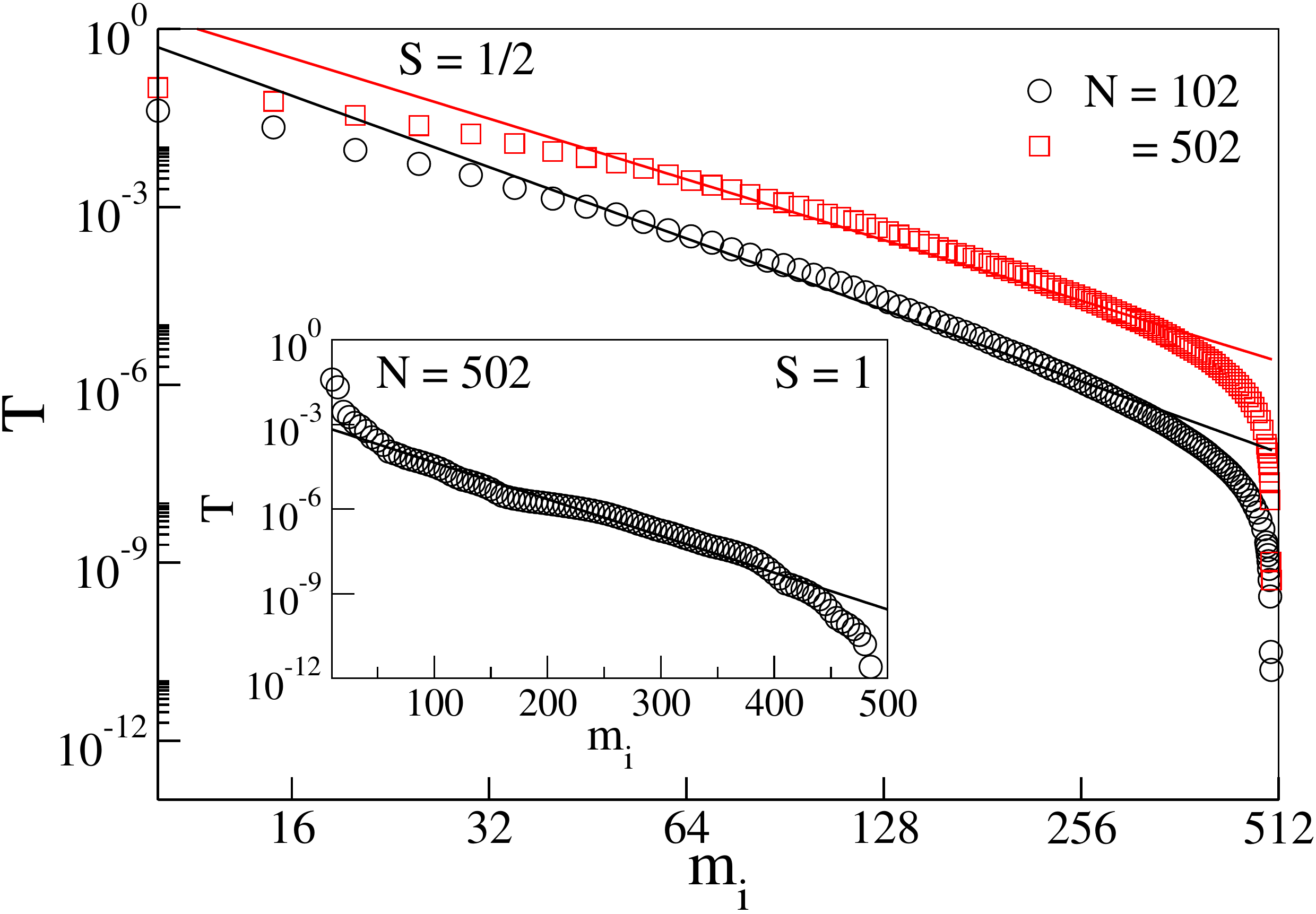}
\caption{\label{fig:trunc}Truncation error $T$ of the density
matrix for spin-1/2 chain (main). The inset shows the truncation error for 
the $S=1$ chain. For $S = 1/2$, the truncation error follows power law decay
whereas it follows exponential decay for a $S=1$ system.}
\end{figure}

As discussed earlier, the DMRG is based on the systematic 
truncation of the irrelevant degrees of freedom and the eigenvalues of 
the density matrix represent the importance of the respective 
states. In the DMRG, only selected states corresponding to the highest 
eigenvalues are kept and the rest of the other states are thrown. We 
define truncation error $T$ as

\begin{align}
T = 1 - \sum_{i=1}^m \lambda_i
\end{align}
where $\lambda_i$ are eigenvalues of density matrix of the 
system block arranged in descending order.   In the main 
Fig.~\ref{fig:trunc}, $T$ is  shown as function of $m$ for 
$S=1/2$ and  the inset shows the same for a $S = 1$ system with N=102  and 502. 
We notice that  $m \sim 350$ for $S=1/2$ and $m \sim 300$ for $S=1$ are sufficient to 
achieve $T=10^{-9}$.  In the main Fig.~\ref{fig:trunc},  $T$ vs $m$ in a log-log 
plot shows a linear behavior,  i.e.,  $T$ for both  system sizes  
of $N=102$ and 502 for $S=1/2$ follows a power law $T \propto m_i^{\alpha}$
with $\alpha = 4.0$ and 3.4, respectively.
The $m$ dependence of $T$  for $S=1$ ring is shown in the inset of 
Fig.~\ref{fig:trunc}.  The truncation error $T$ in this case
decays exponentially, i.e., $T \propto \exp(-\beta m_i)$ with $\beta = 0.03$ for
both $N = 102$ and 502.
\begin{figure}
\centering
\includegraphics[width=\columnwidth]{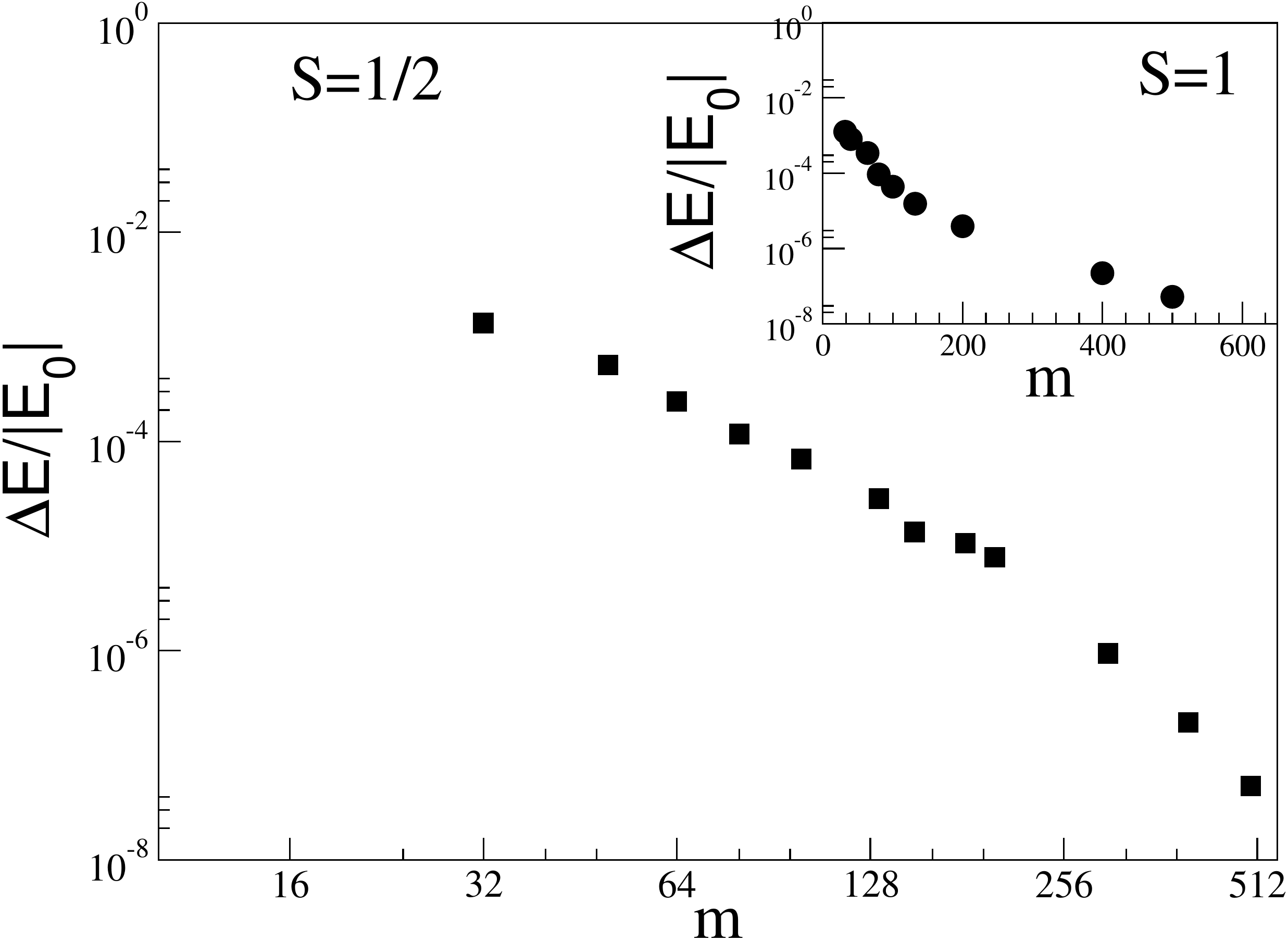}
\caption{\label{fig:nrgacc}Energy accuracy $\Delta E / |E_0|$ 
for spin half chain with PBC (main) which shows a power law 
behavior with $m$. The inset shows the energy accuracy for 
spin one chain with PBC which shows exponential behavior with $m$.}
\end{figure}

The relative error in energies $\Delta E /|E_0|$ for $S=1/2$ and 1 
with $N=102$ spins are shown in Fig.~\ref{fig:nrgacc}, main and inset
respectively. The exact energies of a spin-1 system is 
$E_0/N \sim −1.4014840386$ and this value is obtained by using 
conventional DMRG with $m=2000$ and up to  $N=100$ site chain~\cite{pippan2010}. 
Extrapolation of energies with $m$ is done to obtain the above value~\cite{pippan2010}.  
We notice that $\frac{\Delta E}{E_0}$ for $S=1/2$ systems goes to $10^{-6}$
for $m = 256$ whereas it goes to  $10^{-8}$ for $m \approx 500$, as 
shown in the main Fig.~\ref{fig:nrgacc}. Although similar accuracy 
of the energy can be achieved with $m = 100$ in MPS approach, the 
scaling is $\sim m^4$, rather than $\sim m^3$ in our algorithm.
For $S=1$ accuracy of $10^{-8}$  can be reached with $m \sim 450$,
as shown in the inset of Fig.~\ref{fig:nrgacc}. 
\begin{figure}
\centering
\includegraphics[width=0.93\columnwidth]{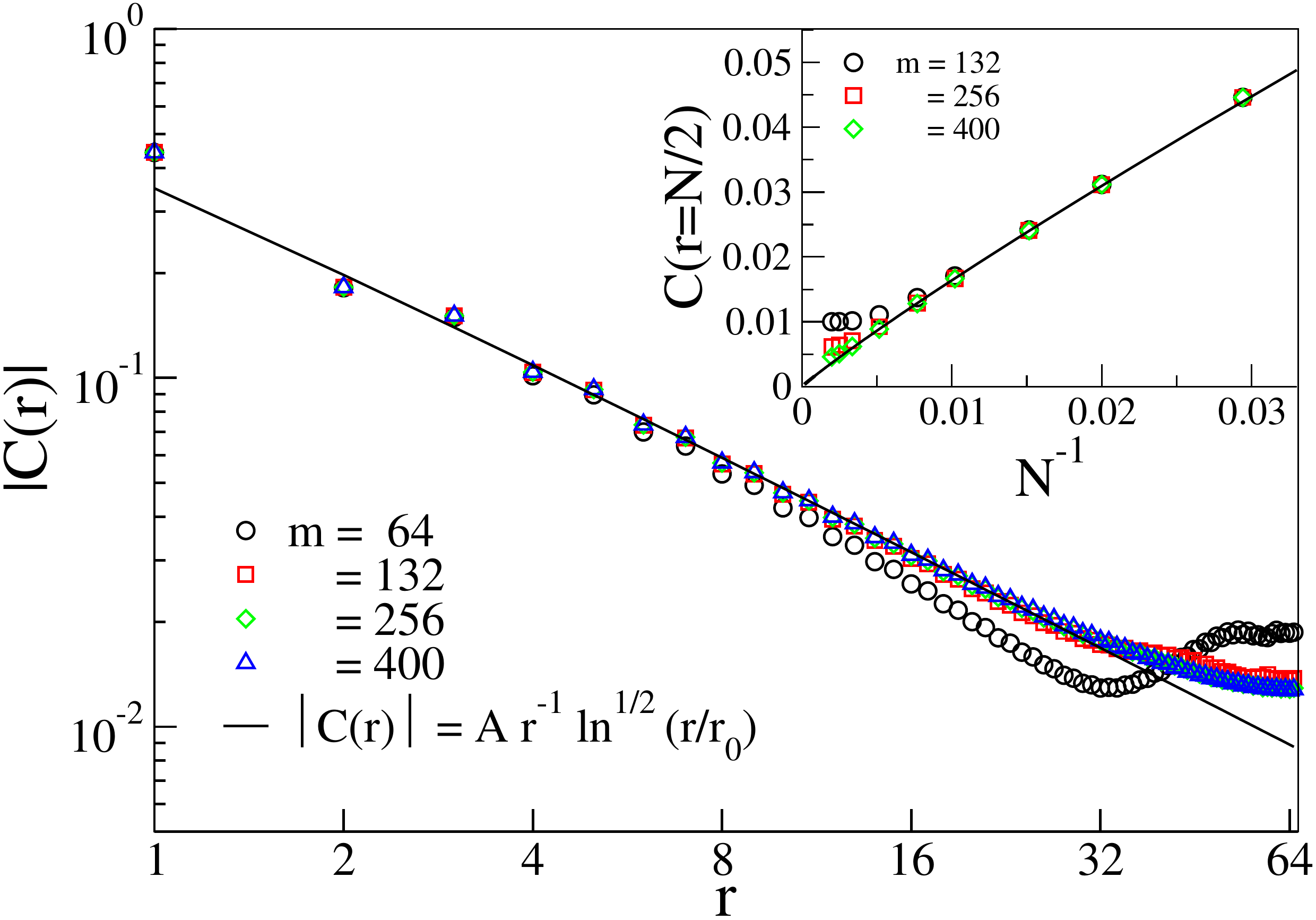}
\caption{\label{fig:rcr} The main figure shows the variation of $|C(r)|$ 
(defined in Eq.~(\ref{eq:crdef})) as a function of $r$ 
for different $m$. The solid curve in the main figure is the logarithmic 
correction formula: $A r^{-1} \ln^{1/2} (r/r_0)$ with $A = 0.22$ and $r_0 = 0.08$.~\cite{sandvik2010,affleck89}
The inset shows the variation  of $C(N/2)$ with the inverse of 
$N$ for different $m$. The solid curve is the form: $A (N/2)^{-1} \ln^{1/2} (N/2r_0)$
with $A = 0.323$ and $r_0 = 0.08$.}
\end{figure}

The dependence of accuracy of correlation function $|C(r)|$ of 
$S=1/2$ for $N=130$ as a function of $m$  is shown in the main Fig.~\ref{fig:rcr}. We 
notice that $m=256$ is sufficient to achieve an accuracy of $\sim 10^{-4}$. 
We also notice that $|C(r)| \propto r^{-1}$ with the well known logarithmic correction
specially for smaller $r$~\cite{affleck89}. 
We have fitted the correlations with the well known form 
$|C(r)| = A r^{-1} \ln^{1/2}(r/r_0)$ with $A = 0.22$ and $r_0 = 0.08$~\cite{sandvik2010}.
Deviation in function for large $r$ is the artifact of finite systems. 
In our algorithm, two sites are 
added symmetrically and the new sites are  added $N/2$ sites apart, 
consequently the distance between two new sites is $N/2$. Therefore, 
the new-new sites correlation function is $C(N/2)$ and is plotted 
with $N^{-1}$ in the inset of Fig~\ref{fig:rcr}. We observe that $m=256$ 
is sufficient for $N \sim 200$ to achieve sufficient numerical accuracy. 
The curve behaves almost linearly with the logarithmic correction: 
$|C(N/2)| = 2A N^{-1}\ln^{1/2} (N/2r_0)$ with $A = 0.323$
and $r_0 = 0.08$.  

\section{\label{sec:sum}Summary}

The DMRG is a state-of-the-art numerical technique for solving the 
1D quantum many-body systems with open boundary condition. 
However, the accuracy of the 1D PBC system is rather poor. 
The MPS approach gives very accurate 
results but the computational cost goes as (o) $m^5$~\cite{verstraete2005}. 
Later, this algorithm was modified and the computational cost of 
the modified  algorithm goes as 
(o) $p \times m^3$, where $p$  in general varies linearly with 
$m$~\cite{rossini2011a}, but $p$ can go as $m^2$ in case of long
range order in the system. The computational cost of the algorithm presented in this paper 
scales as (O) $m^3$ because of the sparse superblock Hamiltonian and is very similar 
to the conventional DMRG. To achieve this goal, we avoid 
the multiple times renormalization of the operators which are
used to construct the superblock. This algorithm can readily be used to solve
general 1D and quasi-1D systems, e.g., $J_1$--$J_2$ model, two-legged ladder. The new algorithm can be implemented with ease
using the conventional DMRG code. 

Our calculation suggests that most of the quantities, e.g., ground state energies, 
energy gaps and correlation function, can accurately be calculated by keeping
$m \sim 400$. The superfluity stiffness~\cite{rossini2011} and dynamical 
structure factors using the correction vector 
technique~\cite{ramasesha-prb99} or continued fraction 
method~\cite{cont-frac-hallberg} can be calculated with this 
algorithm. The symmetries, e.g., spin parity, electron-hole, 
inversion, can easily be implemented in this algorithm~\cite{ramasesha-prb99}.
This algorithm is used by us in calculating accurate 
static structure factors and correlation function for $J_1-J_2$  
model for a spin-1/2 ring geometry~\cite{soos-jpcm-2016}.     

\begin{acknowledgements}
We thank Z. G. Soos for his valuable comments.
M. K. thanks DST for Ramanujan fellowship and computation 
facility provided under the DST project SNB/MK/14-15/137. 
\end{acknowledgements}

\end{document}